\documentclass[a4paper,fleqn,usenatbib,useAMS]{mnras}
 \usepackage{graphicx}	
\usepackage{amsmath}	
\usepackage{amssymb}    
\usepackage{amssymb}	
\usepackage{multicol}	
\usepackage{bm}		
\usepackage{pdflscape}	
\usepackage{hyperref}
\usepackage{subfig}
\usepackage{amsmath}

\usepackage{color}

\newcommand{\fbnd}{f_{\rm da}}
\newcommand{\fvir}{f_{\rm orb}}
\newcommand{\rt}{R_{\rm edge}}
\newcommand{\rlambda}{R_{\rm{\lambda}}}
\newcommand{\zcen}{z_{\rm cen}}
\newcommand{\sigmabnd}{\sigma_{v,\rm da}}
\newcommand{\sigmalos}{\sigma_{v,\rm los}}
\newcommand{\sigmainf}{\sigma_{v,\rm inf}}
\newcommand{\sigmavir}{\sigma_{v,\rm orb}}
\newcommand{\sigmaplos}{\sigma_{\rm p, los}}
\newcommand{\sigmapinf}{\sigma_{\rm p,inf}}
\newcommand{\sigmapvir}{\sigma_{\rm p,orb}}
\newcommand{\zp}{z_{\rm p}}
\newcommand{\lambdap}{\lambda_{\rm p}}

\newcommand{\hMpc}{h^{-1}\ {\rm Mpc}}
\newcommand{\redmapper}{redMaPPer}

\newcommand{\aone}{a_{\rm{1}}}
\newcommand{\atwo}{a_{\rm{2}}}
\newcommand{\bzero}{b_{\rm{0}}}
\newcommand{\bone}{b_{\rm{1}}}
\newcommand{\czero}{c_{\rm{0}}}
\newcommand{\cone}{c_{\rm{1}}}
\newcommand{\ctwo}{c_{\rm{2}}}
\newcommand{\Mpc}{\mbox{Mpc}}

\newcommand{\rtp}{R_{\rm p}}
\newcommand{\avg}[1]{\langle #1 \rangle}

\newcommand{\bound}{orbiting\ }
\newcommand{\bnd}{orb}

\title[The Edge of redMaPPer Clusters]{Clusters Have Edges: The Projected Phase Space Structure of SDSS redMaPPer Clusters}
\author[Tomooka et al.]{Paxton Tomooka$^1$\thanks{E-mail: ptomooka@email.arizona.edu}, Eduardo Rozo$^{1}$, Erika L. Wagoner$^{1}$, Han Aung$^{2}$, 
\newauthor 
Daisuke Nagai$^{2,3}$, Sasha Safonova$^{1,3}$
\\
$^{1}$Department of Physics, University of Arizona, Tucson, AZ 85721, USA \\
$^{2}$Department of Physics, Yale University, New Haven, CT 06520, USA \\
$^{3}$Department of Astronomy, Yale University, New Haven, CT 06520, USA
}

\begin{document}

\maketitle 

\label{firstpage}

\begin{abstract}
We study the distribution of line-of-sight velocities of galaxies in the vicinity of SDSS \redmapper\ galaxy clusters. Based on their velocities, galaxies can be split into two categories: galaxies that are dynamically associated with the cluster, and random line-of-sight projections.  Both the fraction of galaxies associated with the galaxy clusters, and the velocity dispersion of the same, exhibit a sharp feature as a function of radius.  The feature occurs at a radial scale $\rt \approx 2.2\rlambda$, where $\rlambda$ is the cluster radius assigned by \redmapper.  We refer to $\rt$ as the ``edge radius.''  These results are naturally explained by a model that further splits the galaxies dynamically associated with a galaxy cluster into a component of galaxies \bound the halo and an infalling galaxy component.  The edge radius $\rt$ constitutes a true ``cluster edge'', in the sense that no \bound structures exist past this radius.  A companion paper (Aung et al. 2020) tests whether the ``halo edge'' hypothesis holds when investigating the full three-dimensional phase space distribution of dark matter substructures in numerical simulations, and demonstrates that this radius coincides with a suitably defined splashback radius.  
\end{abstract}

\section{Introduction}
\label{intro}

Galaxy clusters are a well known probe of cosmology and galaxy formation \citep[see reviews by][]{Allen2011,kravtsov2012}.  Galaxy clusters are hosted by massive dark matter haloes, so the study of clusters can help us understand both the growth of structure in the Universe, and how galaxies populate haloes in an environment dependent way.  Within the context of galaxy evolution in particular, it is especially important to correctly identify the ``edge'' of a halo/cluster, so that the impact of a galaxy falling into a cluster can be adequately characterized.  Moreover, cosmological inferences may also be sensitive to the choice of boundary adopted when defining haloes/clusters \citep[][Garcia et al. 2020]{garciarozo19}, demonstrating that the adoption of different halo definitions may have cosmological implications as well.

Following early work by \citet{diemerkravtsov14} that showed the outer halo density profile steepens significantly at a characteristic scale, \citet{adhikarieatl14} demonstrated that this steepening can be associated with the splashback radius of the halo: the apocentric radius of particles that have had one passage through the halo.  \citet{moreetal15} proposed that this splashback radius provides a more physical definition of the halo edge, albeit one that depends on the accretion rate of the halo \citep{diemerkravtsov14,adhikarieatl14,diemeretal17}.  Later works have detected a splashback-like feature in the galaxy density profile of photometrically selected \citep{moreetal16,baxteretal17,changetal18} and SZ-selected \citep{shinetal19,zurchermore19} galaxy clusters. 

The splashback feature is traditionally associated with a steepening of the halo density profile.  This has led existing searches for the splashback feature to focus on identifying a ``dip'' in the first derivative of the projected galaxy density profile of galaxy clusters, an inherently difficult and noisy measurement. Here, we investigate whether a similar feature arises in the distribution of line-of-sight velocities of galaxies in the vicinity of a galaxy cluster.  We use the \redmapper\ cluster sample constructed using imaging from the Sloan Digital Sky Survey (SDSS).  We note that a splashback feature in the momentum correlation function has been measured and is even more prominent there than in the halo--mass correlation function \citep{okumuraetal18}.  This establishes the possibility of detecting the splashback feature through galaxy dynamics.

Our analysis reveals the existence of a sharp feature in the velocity distribution of galaxies around \redmapper\ clusters.  We suggest that this feature represents the edge of a halo, in the sense that no \bound galaxies exist beyond the radius we have identified.  In a companion paper (Aung et al. 2020), we demonstrate that this basic conclusion also holds when looking at the three-dimensional phase space distribution of dark matter substructures in numerical simulations.  There, we also establish the relation between the halo edge we have identified, and the splashback radius using particle trajectories as per \citet{diemeretal17}.  The fact that the edge of galaxy clusters can be so easily identified using line-of-sight velocity information should enable a broad range of new studies probing the sensitivity of this halo edge to halo accretion rates, the nature of dark matter, and even modified gravity theories \citep[e.g.][]{diemeretal17,adhikarietal18,adhikarietal19}.

Unless otherwise noted, all cosmology dependent quantities in this work were calculated assuming a flat $\Lambda$CDM model with $\Omega_{\rm m}=0.3$ and $H_0=100h\ \mbox{km/s/Mpc}$ (i.e. we work in $h^{-1}\ \Mpc$ units).

\section{Data}
\label{data}

Our analysis requires both a spectroscopic galaxy catalog, and a cluster catalog.  Our spectroscopic data set is comprised of all spectroscopic galaxies released as part of the SDSS DR14 \citep{dr14}.  This constitutes a total of $\sim 2.6$M spectroscopic galaxies across $\approx 10,000\ \deg^2$ of the northern and southern sky.  We determine the redshifts of the central galaxies of clusters, as well as the velocities of neighboring galaxies relative to the central galaxies, using SDSS spectroscopic redshifts.  

The cluster catalog we use in our analysis is the SDSS DR8 \redmapper\ cluster catalog, v5.10 \citep{rmIV}.  \redmapper\ is a red-sequence cluster finding algorithm \citep{rmI} that iteratively self-trains the model for red-sequence galaxies as part of the cluster finding.  \redmapper\ cluster catalogs are both pure and complete, and the optical richness is a good proxy for cluster mass \citep{rmII, rmIII,simetetal17}.  We restrict our work to the publicly available cluster catalog, with a richness threshold $\lambda \geq 20$.

In this work, we wish to study the phase space structure of \redmapper\ clusters.  To that end, we restrict ourselves to \redmapper\ clusters whose central galaxy has a spectroscopic redshift. This reduces the full \redmapper\ catalog of $\sim 27$k clusters to $\sim 17$k systems with a spectroscopic central galaxy.  We adopt this spectroscopic redshift as the redshift of the galaxy cluster.  Because the \redmapper\ catalog did not store the spectroscopic redshift uncertainties, we cross-match the \redmapper\ catalog to the SDSS DR14 spectroscopic catalog using a 2~arcsec aperture.  Clusters that have an apparent match but a poor redshift match ($|\Delta z| \geq 0.03$) are discarded (0.03\% of the sample). 

At high redshfits, SDSS photometry is not sufficiently deep to detect all cluster galaxies contributing to the richness definition adopted by \redmapper.  Due to this incompleteness, we limit ourselves to a volume-limited sub-sample of the \redmapper\ catalog defined by the spectroscopic redshift cuts $z\in[0.1,0.3]$, which reduces the number of clusters to $5,015$ systems.  A similar (photometrically defined) volume-limited sub-sample has been used to place cosmological constraints \citep{costanzietal18b,kirbyetal19}.


\section{Measuring The Phase Space Structure of redMaPPer Clusters}
\label{two population}

We measure the phase space structure of \redmapper\ clusters by stacking velocity histograms as done, for example, in \citet{rmIV} and \citet{farahietal16}.  Critically, however, we will stack the velocity data using galaxies within narrow radial bins, allowing us to measure the radial dependence of the velocity distribution of galaxies in the vicinity of galaxy clusters.

\subsection{Identifying Potential Central--Satellite Pairs}
\label{create catalog}

To probe the phase space structure of SDSS \redmapper\ clusters, we collect all spectroscopic galaxies within a $5\rlambda$ radius of each galaxy cluster in our volume- and richness-limited spectroscopic cluster sample.  \redmapper\ estimates cluster richness by counting galaxies using the richness-dependent radius $\rlambda$.  By definition \citep{rmI}, this radius is related to the richness via
\begin{equation}
    \rlambda = (1\ \hMpc) \left( \frac{\lambda}{100} \right) ^{0.2}
\end{equation}
Thus, to a rough approximation, we select all galaxies within a $5\ \hMpc$ radius of each of our galaxy clusters.

Next, for each central-satellite pair, we compute the line-of-sight velocity of the member relative to the cluster as
\begin{equation}
    v = c\frac{z_{\rm sat}-\zcen}{1+\zcen}.
\label{calculate velocity}
\end{equation}
Plotting the line-of-sight velocity $v$ against richness clearly reveals two populations (see e.g. Figure~2 in \citet{rmIV}): a set of galaxies with velocities of order $\approx 10^3\ \mbox{km/s}$ that are clearly associated with the cluster, and a set of non-cluster members with much larger apparent velocities. These large relative velocities are due to galaxies at very large distances from each other along the line-of-sight.  To reduce this contamination, we make a conservative by-eye cut that rejects most of the unassociated galaxies along the line-of-sight while still preserving all the cluster galaxies associated with the galaxy clusters.  Following \citet{rmIV}, the cut we applied is
\begin{equation}
    |v| \leq  (3000\ \text{km/s})(\lambda/20)^{0.45}.
\label{velocity cut}
\end{equation}
To ensure that the redshift of nearby galaxies is unaffected by light from the central galaxy, we also impose a small scale radial cut at $R\geq 0.05\ \hMpc$.  Nearby galaxies that project onto a smaller radial distance than this are discarded.  Finally, we trim all central-satellite pairs for which either galaxy has a DR14 redshift error $>10^{-4}$. This ensures that the spectroscopic redshift uncertainties in the velocity are small ($\Delta v \sim 30~\mbox{km/s}$) relative to the velocity dispersion of the galaxy clusters.  The fraction of potential spectroscopic central--satellite pairs discarded because of the reported redshift error is 2.2\%.  The final number of potential central--satellite pairs is $\sim 87$k.


\subsection{The Measurement}
\label{two population methods}

Having identified all potential central--satellite galaxy pairs, we separate the galaxies into radial bins.  To roughly account for the broad richness distribution of clusters, we define our radial bins in terms of the radius $R$ measured in units of the cluster radius $\rlambda$.  The specific radial bins we consider are $R/\rlambda\in (0.0,0.2], (0.2,0.4], (0.4,0.6]\ldots(4.6,4.8], (4.8,5.0]$.

\begin{figure*}
    \centering
    \includegraphics[width=0.95\textwidth]{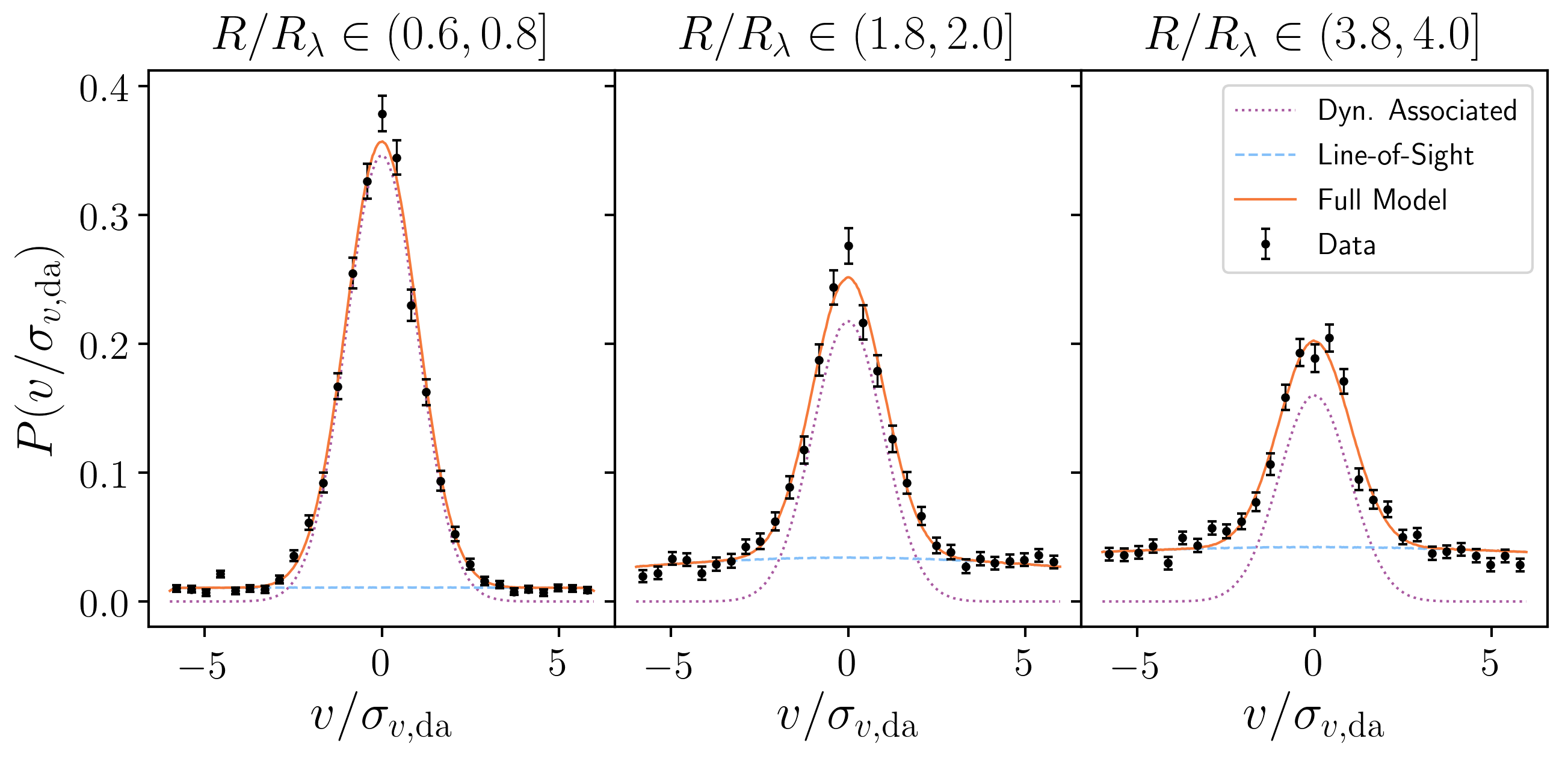}
    \caption{Histograms of the line-of-sight velocity of galaxies in the vicinity of a galaxy cluster, relative to the central galaxy of the cluster.  The velocity of each galaxy has been rescaled by the best-fit velocity dispersion $\sigmabnd$ for that galaxy, where the latter is determined as a function of the richness and redshift of the galaxy cluster hosting the central--satellite galaxy pair.  Each of the three panels corresponds to galaxies in different radial bins, as labelled. The histogram is modeled as a sum of two galaxy populations: a population of dynamically associated galaxies (purple dotted), and random line-of-sight projections (blue dashed). The sum of these two gives rise to the full model (orange solid).  As discussed in the paper, this simple two-component model will be superseded by a more physical model described in detail in section~\ref{threepopulation}. 
    }
    \label{fig:histo1}
\end{figure*}

We construct velocity histograms for all radial bins, and find that they are all qualitatively similar: there is a large roughly Gaussian peak due to galaxies associated with the cluster, and a broad ``shelf'' of galaxies due to uncorrelated structure along the line-of-sight (see Figure~\ref{fig:histo1}).  At each radial bin, we model the peak using a Gaussian distribution with zero mean velocity.  The standard deviation of the distribution is modeled as a power-law in richness and redshift, 
\begin{equation}
    \sigmabnd(\lambda, \zcen) = \sigma_p \left (\frac{1+\zcen}{1+\zp} \right )^\beta \left(\frac{\lambda}{\lambdap} \right)^\alpha
\label{vd}
\end{equation}
where $\zp$ and $\lambdap$ are the redshift and richness pivot points.  We choose these to be the median redshift ($\zp = 0.172$) and median richness ($\lambdap = 32.192$) of the central--satellite pairs. For each radial bin, we fit for the three parameters describing the velocity distribution of galaxies in that radial bin. The parameters are: $\sigma_p$ (pivot velocity dispersion), $\alpha$ (richness scaling), and $\beta$ (redshift evolution).  In addition, we describe the line-of-sight contamination as a much wider Gaussian with mean zero and standard deviation $\sigmalos$.  We explicitly account for the velocity cut applied in our galaxy selection by truncating the model distribution at the applied velocity cut.  This leads us to renormalize the Gaussian distribution describing the line-of-sight contaminants with the appropriate error function.


The full likelihood of observing a galaxy pair of line-of-sight velocity $v$ is given by 
\begin{equation}
    \mathcal{L}_i = \fbnd G(v_i|\sigmabnd) + (1-\fbnd)G(v_i|\sigmalos)
\label{two_pop_li}
\end{equation}
where $\fbnd$ is the fraction of galaxies dynamically associated with the galaxy clusters (hence the subscript ``da'').  Altogether, \it for each radial bin \rm this model has five free parameters: $\sigma_p$, $\alpha$, and $\beta$ from the velocity dispersion of dynamically associated galaxies, $\sigmalos$ from the line-of-sight noise, and $\fbnd$ which normalizes this two-population model to unity. The full likelihood for the data set is the product of the individual likelihoods for galaxies within a single radial bin,
\begin{equation}
    \mathcal{L} = \prod_i \mathcal{L}_i.
\label{full_like}
\end{equation}
We use a Markov chain Monte Carlo (MCMC) to determine the posterior distribution for each of our model parameters in each of our radial bins.

It is worth nothing that we do not believe fiber collisions have any impact on our results.  Velocity measurements are clearly insensitive to fiber collisions: fiber collisions may lead to missing data, but they don't bias the velocity measurements we obtain.  Obviously the spatial distribution of the cluster pairs will be impacted, but that is not a statistic with which we are concerning ourselves. 


\subsection{Results}
\label{results}


\begin{figure*}
\centering
\includegraphics[width=0.47\textwidth]{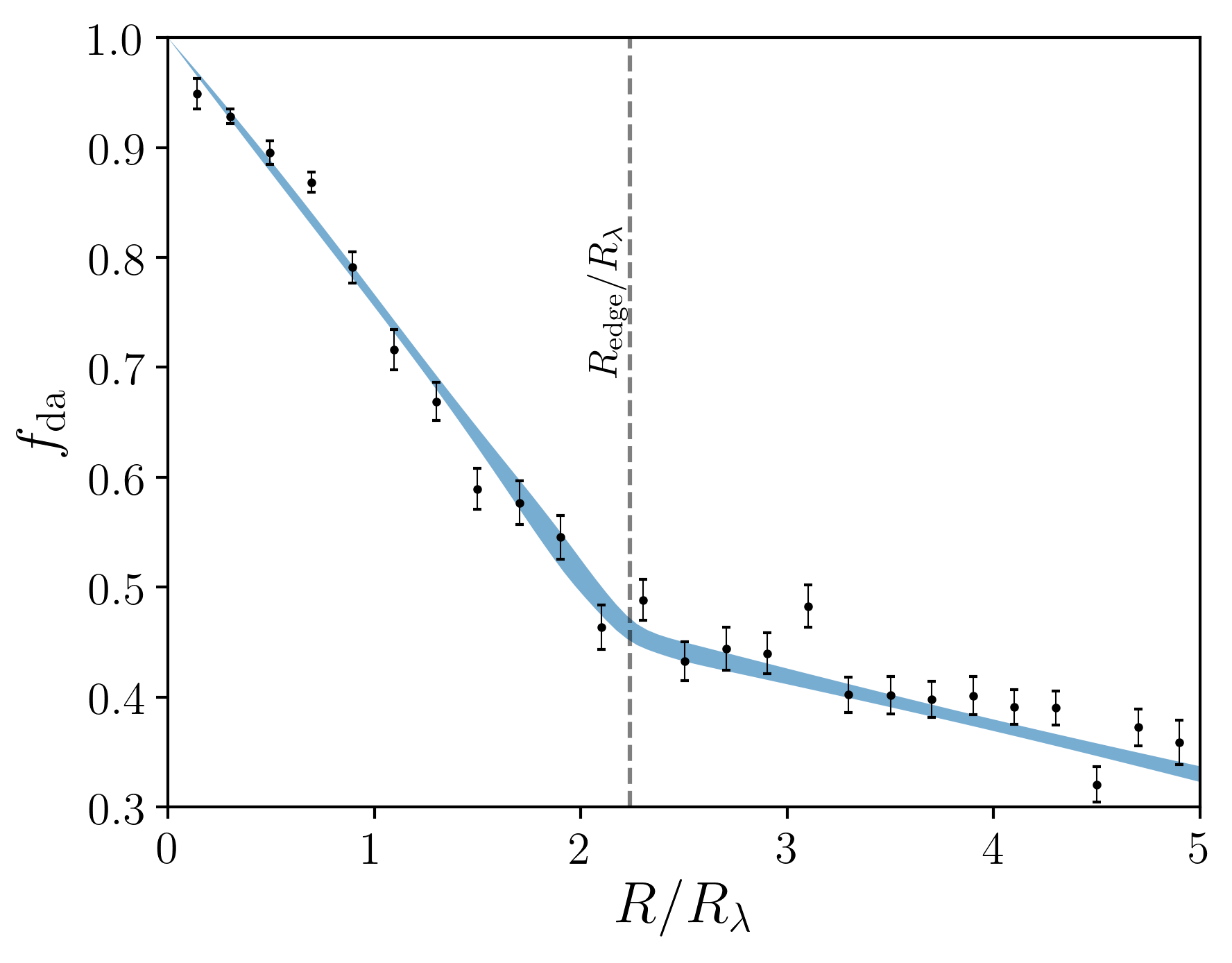}
\includegraphics[width=0.47\textwidth]{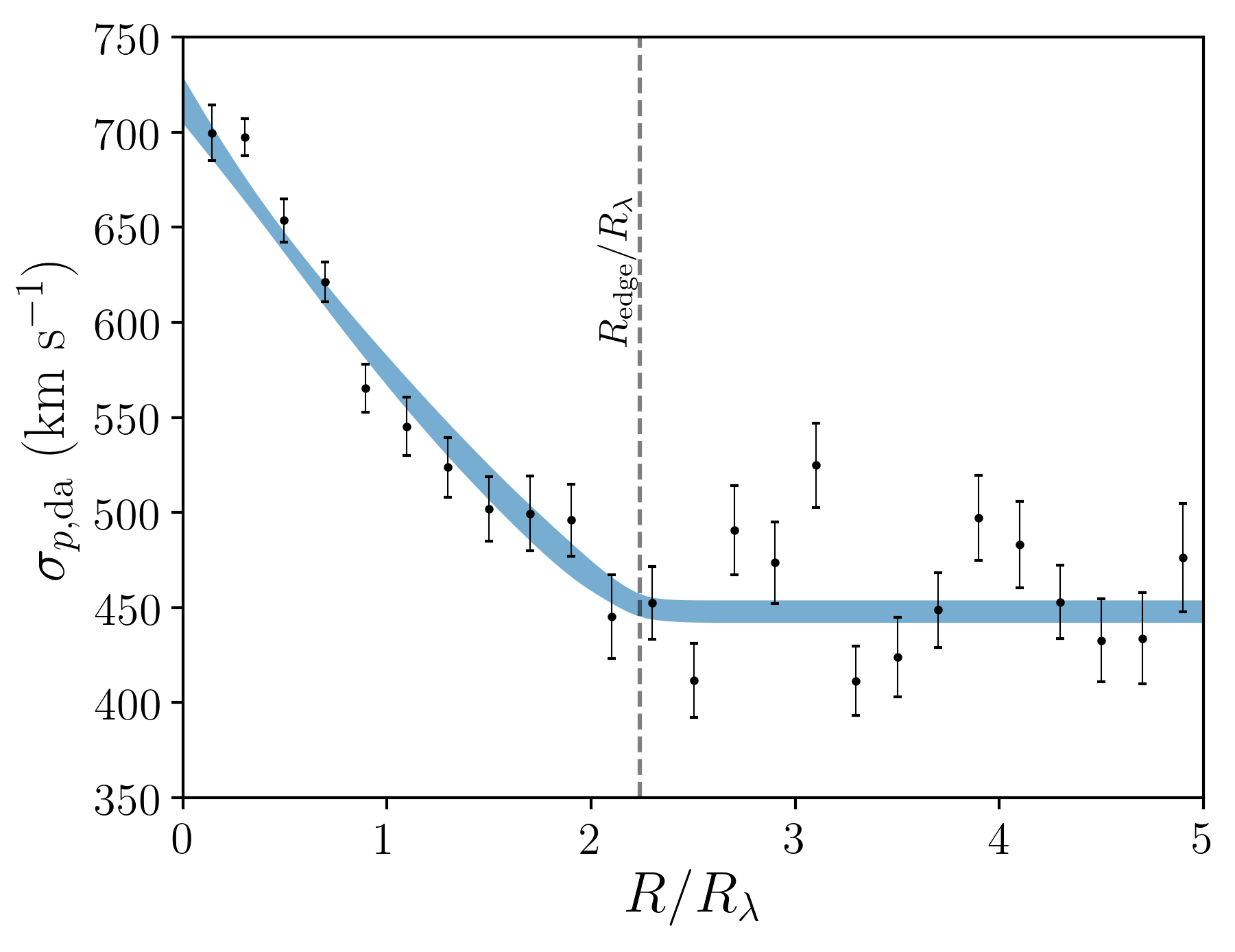}
\caption{{\bf Left:} Fraction of galaxies $\fbnd$ dynamically associated with the galaxy clusters as a function of cluster radius.  The points with error bars correspond to the measurements in each individual radial bin.  The blue band shows the 68\% region of the posterior from our final model detailed in section~\ref{threemodel}. The edge radius, marked by the vertical dashed line, is taken from the fit of our final model.  {\bf Right:} The velocity dispersion $\sigmabnd$ of the galaxies dynamically associated with \redmapper\ clusters as a function of radius. Remarkably, the velocity dispersion appears to be constant beyond the edge radius $R/\rlambda\approx 2.2$.}
\label{fig:fbound}
\end{figure*}


Figure \ref{fig:fbound} shows the recovered parameters $\fbnd$ (left) and $\sigmabnd$ (right) as a function of the cluster radius $R/\rlambda$. Some of the basic trends are easy to understand: the fraction of galaxies dynamically associated with the galaxy cluster decreases as a function of radius.  Likewise, starting from $R=0$, the velocity dispersion of the galaxy cluster decreases with increasing radius, at least until $R/\rlambda\approx 2.2$.  This brings us to the most remarkable feature in these plots: both $\fbnd$ and $\sigmabnd$ exhibit a clear transition at the same cluster radius, roughly $R/\rlambda = 2.2$.  Understanding the origin of this ``knee'' is the focus of the remainder of this paper.  By contrast, both $\alpha$ and $\beta$ are roughly constant as a function of cluster radius (not shown).  For future reference, we will denote the radius of this knee as $\rt$, and refer to it as an ``edge radius.''

Figure~\ref{fig:fbound} raises several interesting questions. What is the physical significance of the ``knee'' in the plots? What are the differences between galaxies inside this transition scale and those beyond that transition?  Why is the velocity dispersion of galaxies far from the cluster center apparently independent of cluster radius?
 
We interpret this transition as
a physical barrier that distinguishes between two satellite populations; namely, cluster galaxies orbiting the galaxy cluster, and neighboring galaxies falling into the cluster.  That is, this transition corresponds to a real cluster ``edge.'' Consider the physics governing satellite dynamics as we move from the cluster center outwards. Surrounding the cluster center, there exists a population of cluster galaxies orbiting the halo (subscript ``\bnd'').  By contrast, when we are far from the cluster center, none of the observed galaxies can reasonably be called cluster galaxies: they are simply too far away from the cluster center.   The dynamically associated galaxies we see must instead represent infalling galaxies (subscript ``inf'').  These infall regions extend a large distance along the line-of-sight and contribute to the nearly constant apparent velocity dispersion at large radii as shown in the right panel of Figure~\ref{fig:fbound}.

The fact that there is a sharp transition in Figure~\ref{fig:fbound} strongly suggests that it is at this radius that the \bound galaxies ``turn on,'' so to speak.  That is,  the radius $\rt$ is the maximum radius at which we can find \bound cluster galaxies.  We formalize this idea in the next section, and use it to describe the phase space structure of cluster galaxies across all radii simultaneously.


\section{Characterizing the Phase Space Structure of redMaPPer Clusters}
\label{threepopulation}

\subsection{Model}
\label{threemodel}

Motivated by our discussion in section \ref{results}, we attempt to describe the line-of-sight velocity data across all radial bins simultaneously.  The fundamental insight of our model is that there are three distinct galaxy populations that we need to account for: an orbiting component, an infalling component, and, finally, uncorrelated galaxies along the line-of-sight.  Thus, the likelihood for any one central-satellite pair is given by
\begin{align}
    \mathcal{L}_i & = \fbnd [\fvir G(v_i|\sigmavir) + (1-\fvir) G(v_i|\sigmainf)] \nonumber \\
    & \hspace{1.5in} + (1-\fbnd)G(v_i|\sigmalos)
\label{final_li}
\end{align}
where $\fbnd$ describes the fraction of galaxies dynamically associated with the galaxy cluster, and $\sigmalos$ describes the population of unassociated galaxies along the line-of-sight.  The velocity dispersion of the physically unassociated galaxies $\sigmalos$ is \it not a function of radius. \rm  Of course, the same is not true of $\fbnd$, which is necessarily a decreasing function of radius.  We will return to the model for $\fbnd(R)$ momentarily.  Much like the $\fbnd$ parameter, $\fvir$ serves to scale the amplitudes of the orbiting and infalling galaxies, and is necessarily a function of radius.  In this model, for any given radius, the probability that a galaxy is a \bound or an infalling galaxy is given by
\begin{eqnarray}
P({\rm \bnd}) & = & \fbnd  \fvir \\
P({\rm inf}) & = & \fbnd  (1-\fvir) \\
P({\rm los}) & = & 1-\fbnd
\end{eqnarray}
These probabilities sum to unity.  As before, each of the Gaussians is centered at zero, but we now allow all three velocity dispersions ($\sigma$) to be functions of redshift and richness in the same manner as equation \ref{vd}. We have then 
\begin{eqnarray}
\sigmavir (\lambda , \zcen) & = & \sigmapvir \left (\frac{1+\zcen}{1+\zp} \right )^{\beta_{\rm \bnd}} \left(\frac{\lambda}{\lambdap} \right)^{\alpha_{\rm \bnd}} \label{eq:sigvir1} \\
\sigmainf (\lambda , \zcen) & = &\sigmapinf \left (\frac{1+\zcen}{1+\zp} \right )^{\beta_{\rm inf}} \left(\frac{\lambda}{\lambdap} \right)^{\alpha_{\rm inf}} \\
\sigmalos (\lambda , \zcen) & = &\sigmaplos \left (\frac{1+\zcen}{1+\zp} \right )^{\beta_{\rm los}} \left(\frac{\lambda}{\lambdap} \right)^{\alpha_{\rm los}} \label{eq:siglos}
\end{eqnarray}
contributing a total of nine parameters to our model.  Importantly, only the orbiting velocity dispersion $\sigmavir$ is allowed to vary as a function of radius, as described below.  That radial dependence \it must \rm be there: from the virial theorem, and the fact that the density profile of a cluster increases with decreasing radius, we know that galaxies orbiting at small radii must move faster than galaxies orbiting far from the cluster center.  By contrast, $\sigmainf$ and $\sigmalos$ are assumed to be radius independent.

One may also wonder why the line-of-sight velocity dispersion is allowed to scale with richness and redshift.  The reason is that photometric selection effects may well impact this line-of-sight component relative to a purely mass-selected cluster sample.  Moreover, such a selection bias would almost certainly be richness and redshift dependent, giving rise to the richness and redshift scalings introduced in equation~\ref{eq:siglos}.

Let us turn now to describing the radial dependence of the quantities $\fbnd$, $\fvir$, and $\sigmavir$.  We begin with our description of our $\fbnd$ and $\fvir$. From inspection of Figure~\ref{fig:fbound}, we observe a slightly concave-down, steep declination of $\fbnd$ as we move towards $\rt$, followed by a more gentle, linear decline beyond $\rt$. Based on this observation, we adopt the following model for $\fbnd(R)$,
\begin{align}
    \fbnd(R) & = 
    \begin{cases}
    1+\aone(R/\rt)+\atwo(R/\rt)^2 & \text{for}\ R \leq \rt \\
    \bzero+\bone(R/\rt-1)          & \text{for}\ R \geq \rt .
    \end{cases}
\end{align}
Since the function $\fbnd(R)$ must be continuous at $\rt$, we have the constraint equation $\bzero = 1+\aone+\atwo$. Note we have also demanded that $\fbnd(0)=1$, that is, along the cluster center, all the galaxies we see are dynamically associated with the galaxy cluster.  It is clear from Figure~\ref{fig:fbound} that this is an excellent approximation.   A similar reasoning leads us to model $\fvir$ as
\begin{align}
    \fvir(R) & = 
    \begin{cases}
    \czero+\cone(R/\rt)+\ctwo(R/\rt)^2 & \text{for}\  R \leq \rt \\
    0                        & \text{for}\  R \geq \rt
    \end{cases}
\end{align}
Again, we insist that the function is continuous at $\rt$ which introduces the constraint equation $\czero+\cone+\ctwo=0$.

As mentioned earlier, the velocity dispersion of orbiting galaxies must include a radial dependence: galaxies at small radii must move faster.  The simplest possible model for such a dependence is a linear or possibly power-law dependence on $R$.  Using our intuition that variance is a more primitive quantity than standard deviation, and assuming clusters are self-similar, we consider two possible models, $\sigmavir^2 = \sigmapvir^2(1 - k(R/\rt))$ and $\sigmavir^2 = \sigmapvir^2/(1+k(R/\rt))$.  Here, $k$ is a constant that determines how quickly the velocity dispersion changes as a function of projected cluster radius.  We find that the second option ---  $\sigmavir^2 = \sigmapvir^2/(1+k(R/\rt))$ --- provides a much better fit in simulated data, leading us to adopt the model
\begin{equation}
    \sigmavir (R|\lambda , \zcen) = \frac{\sigmapvir}{\sqrt{1+k\frac{R}{\rt}}} \left (\frac{1+\zcen}{1+\zp} \right )^{\beta_{\rm \bnd}} \left(\frac{\lambda}{\lambdap} \right)^{\alpha_{\rm \bnd}} .
\end{equation}
Again, the key assumption here has been self-similarity of the clusters when physical quantities are plotted in units of $R/\rt$.  The sole remaining ingredient of our model is $\rt$, which we assume to be both richness and redshift dependent.  We set
\begin{equation}
    \rt (\lambda , \zcen) = \rtp \left(\frac{\lambda}{\lambdap} \right)^{\alpha_{\rt}}\left( \frac{1+\zcen}{1+\zp} \right)^{\beta_{\rt}}.
\end{equation}
A by-eye inspection of Figure~\ref{fig:fbound} reveals an edge radius of roughly $2.2 \rlambda$, which at a pivot richness of $\lambda_{\rm p} \approx 32$, equates to $\rt = 1.75\ \hMpc$. Therefore, we expect to find $\rtp\approx 1.75\ \hMpc$ upon fitting our model to the velocity data.

Altogether, our model has 18 parameters: five free variables from $\fbnd$ and $\fvir$ ($a_1$, $a_2$, $b_1$, $c_1$ and $c_2$); four power-law amplitudes describing the richness/redshift dependence of the velocity dispersion of each of the three populations of galaxies we considered, as well as the richness/redshift dependence of the edge radius ($\sigmapvir$, $\sigmapinf$, $\sigmaplos$, and $\rtp$); eight slopes describing the richness and redshift dependence of the velocity dispersion and edge radius (the $\alpha$'s and $\beta$'s); and finally, the $k$ parameter that characterizes the radial dependence of $\sigmavir$. 

While this model may seem complicated, it is, in fact, conceptually simple. We are simply positing that: A) Galaxies can be orbiting the central halo, infalling into the central halo, or be line-of-sight projections,  B) The edge radius $\rt$ is a physical boundary for the halo, and C) That halos are self-similar in $R/\rt$, and the fraction of dynamically associated and orbiting galaxies is radially decreasing.  Everything else follows from these assumptions.  Moreover, while 18 parameters may seem like a lot, 
it is worth noting that the apparently ``simpler'' model of section~\ref{two population methods} required 125 parameters to describe the velocity distribution across all radial bins.  Evidently, the model described in this section is a dramatic simplification which nevertheless captures all of the relevant physical insights garnered from Figure~\ref{fig:fbound}.  Table~\ref{tab:params} summarizes our model parameters, and their posteriors.  All parameters had flat priors, except for $a_1$ and $c_1$, the linear terms of $f_{\rm da}$ and  $f_{\rm orb}$, which we demanded were negative, i.e. the fractions of dynamically associated and orbiting galaxies decrease with radius at zero radius. 


\begin{figure*}
    \centering
    \includegraphics[width=0.9\textwidth]{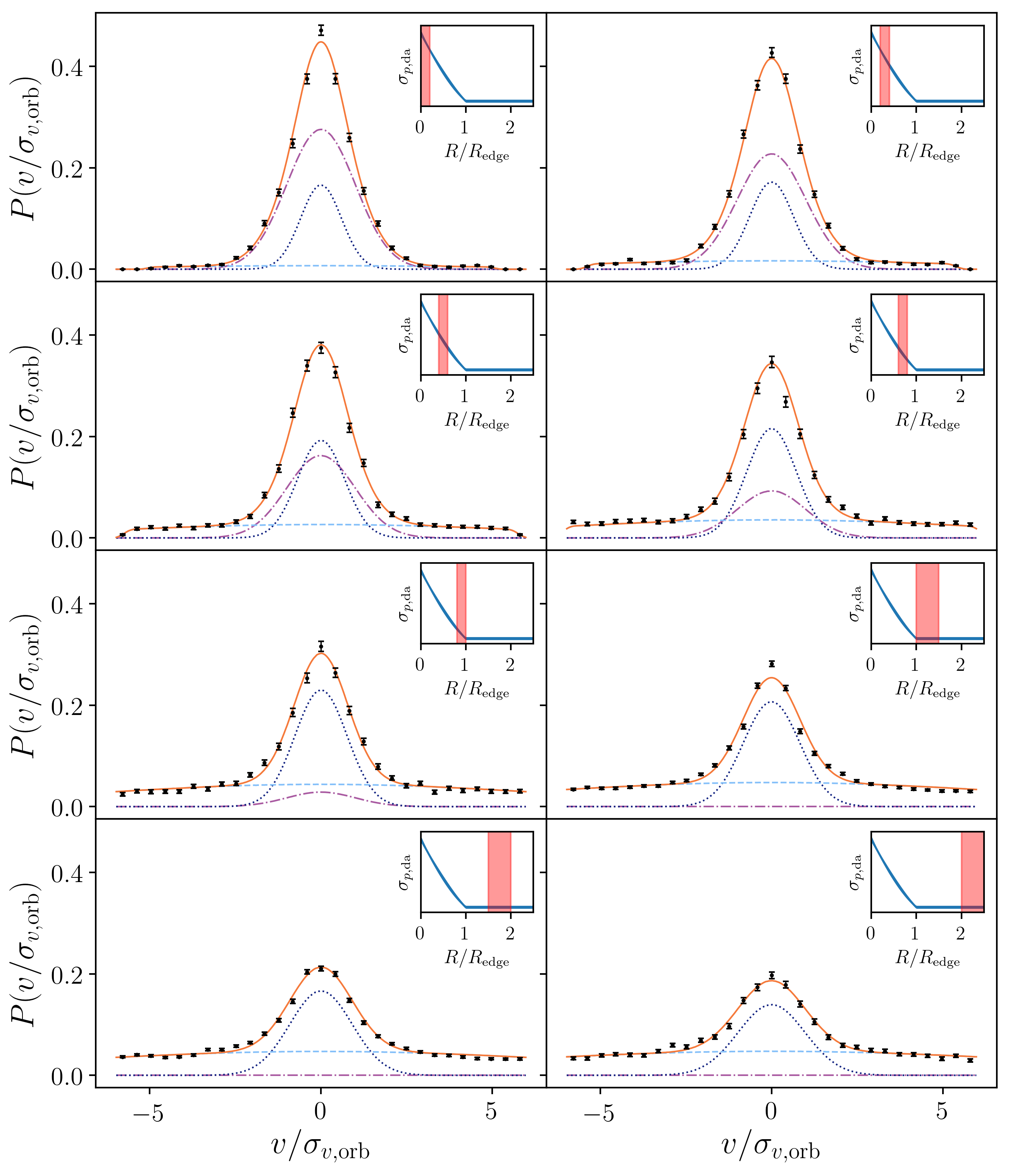}
    \caption{The distribution of line-of-sight velocities of galaxies around \redmapper\ clusters. The points with error bars correspond to the velocity histogram measurements, while the orange solid line is our best fit model.  The remaining three lines correspond to the orbiting galaxy contribution (purple dot-dash), the infalling galaxy contribution (dark blue dotted), and the line-of-sight contribution (light blue dashed).  Each panel is a slice of $R/\rt$, as illustrated by the inset panel.}
    \label{fig:histograms}
\end{figure*}



\subsection{Results}
\label{final_results}


\begin{table*}
  \caption{Model parameters describing the radius-dependent distribution of line-of-sight velocities of galaxies in the vicinity of a SDSS \redmapper\ clusters.  The reported values with errors are the posteriors from our analysis, in the units described below (where appropriate).  In all cases, the subcripts ``orb'', ``inf'', and ``los'' refer to orbiting, infalling, and line-of-sight galaxies. $\fbnd$ is the fraction of galaxies dynamically associated with a galaxy clusters, and $\fvir$ is the fraction of orbiting galaxies.  All parameters had flat priors, except for $a_1$ and $c_1$, the linear terms of $f_{\rm da}$ and  $f_{\rm orb}$, which we demanded were negative, i.e. the fractions of dynamically associated and orbiting galaxies decrease with radius at zero radius. }
  \label{tab:params}
    \begin{tabular}{llll}
		$a_1 = -0.512 \pm 0.065$, &$a_2 = -0.032 \pm  0.053$, &$b_1=-0.010\pm 0.008$ & Radial dependence of $\fbnd$. \\
		$c_1 = -0.061\pm 0.180$, &$c_2 =  -0.693\pm 0.155$ && Radial dependence of $\fvir$. \\
		$\sigmapvir = 7.77 \pm 0.21 $, &$\sigmapinf = 4.48\pm 0.06$, &$\sigmaplos = 38.73\pm 1.16$ & Pivot velocity dispersion in units of 100~km/s. \\
		$\alpha_{\rm orb} = 0.430 \pm 0.019$, &$\alpha_{\rm inf} =  0.363\pm 0.020$, &$\alpha_{\rm los} = 0.503\pm 0.066$ & Richness scaling index of the galaxy velocity dispersion. \\
		$\beta_{\rm orb} = 0.353 \pm 0.195$, &$\beta_{\rm inf} = 0.113\pm 0.180$, &$\beta_{\rm los} = -0.328\pm  0.480$ & Redshift evolution index of the galaxy velocity dispersion. \\
		$\alpha_{\rt} = 0.305 \pm 0.034$, &$\beta_{\rt} = -0.385 \pm 0.289 $ &$\rtp = 1.79 \pm 0.12$, & Edge radius slopes and amplitude (in $\hMpc$). \\
		$k= 0.883 \pm 0.238$ &&& Radial scaling of $\sigmavir$.
    \end{tabular}
\end{table*}


\begin{figure*}
    \centering
    \includegraphics[width=0.49\textwidth]{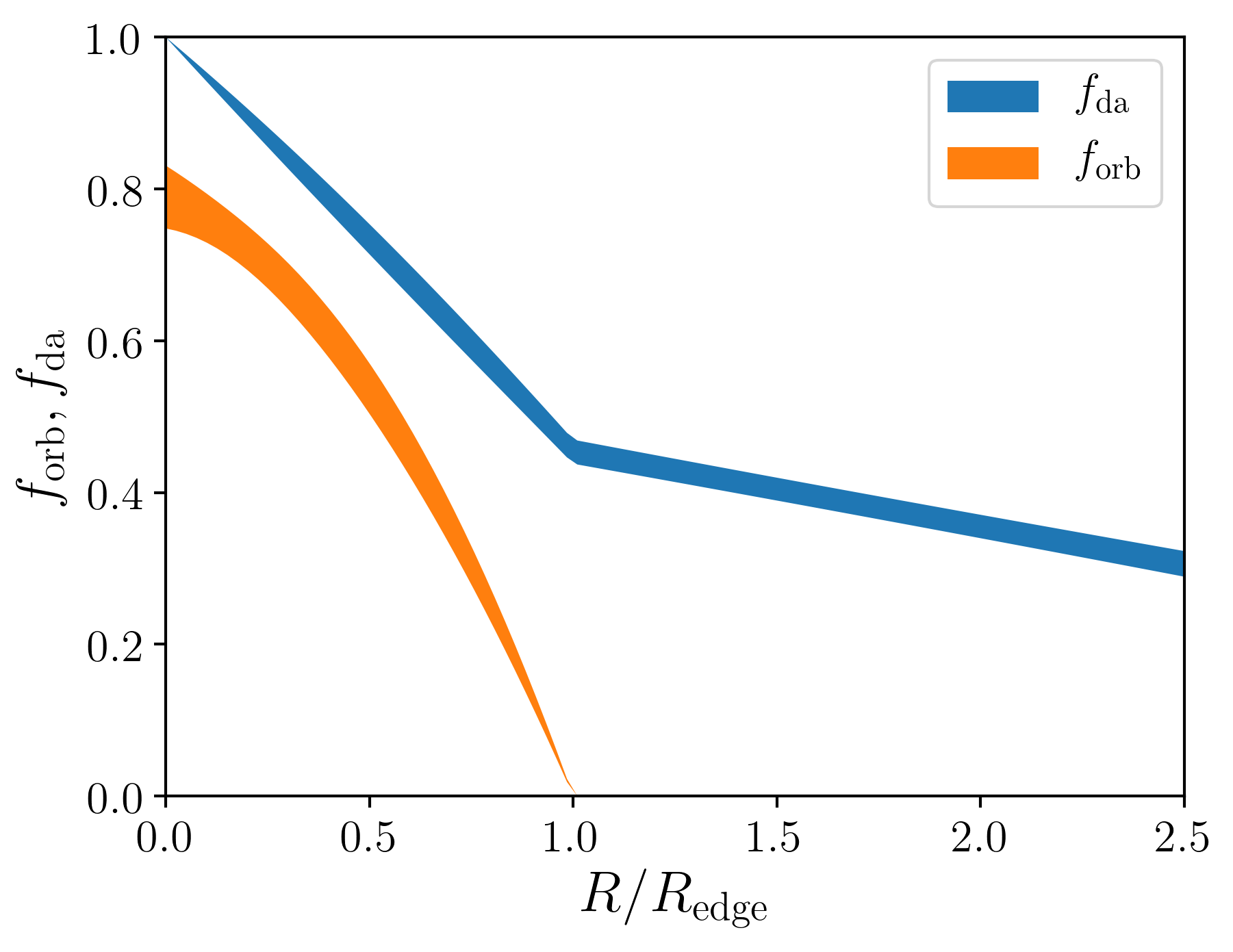}
    \includegraphics[width=0.49\textwidth]{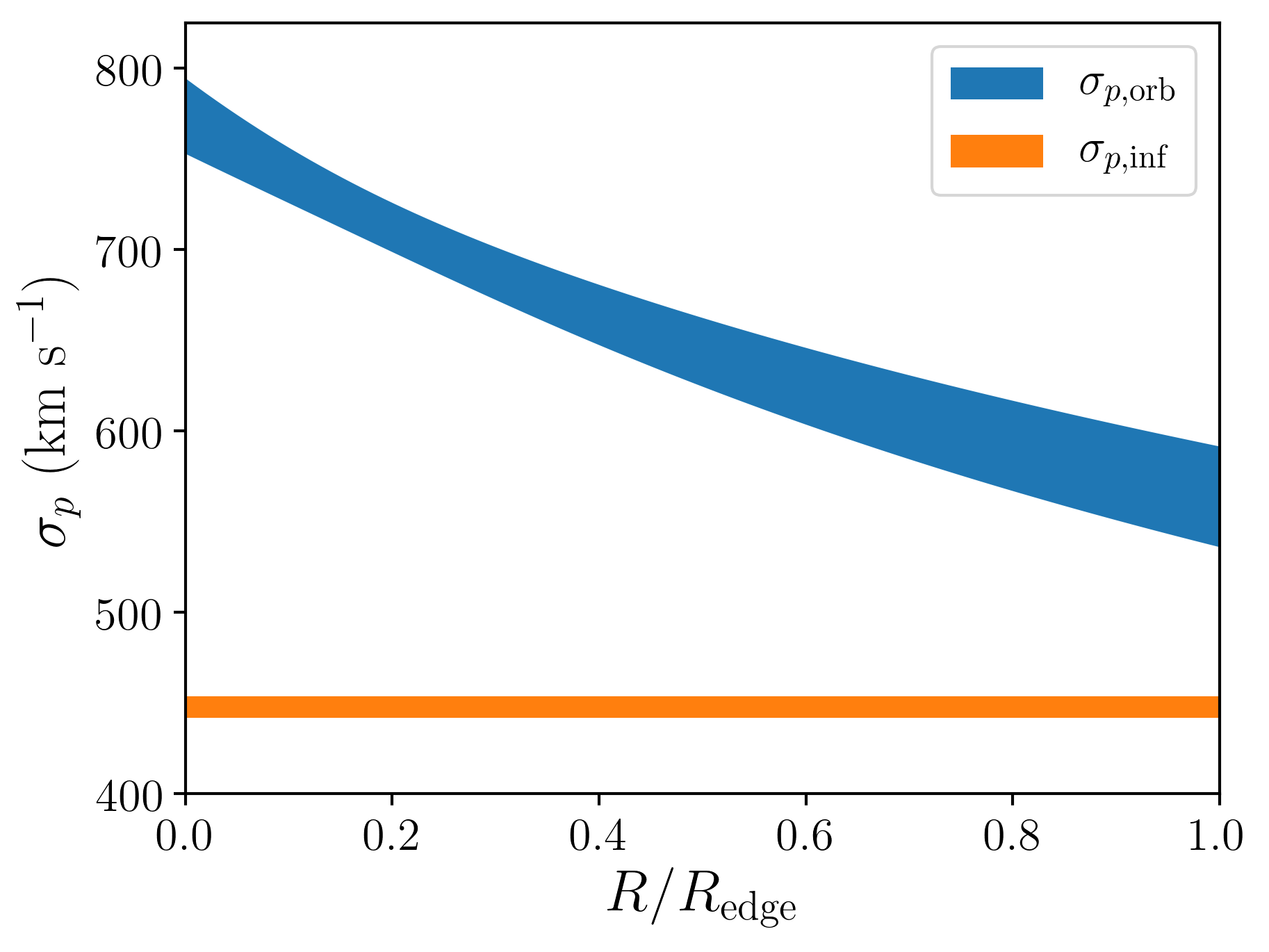}
    \caption{The two panels above show a graphical summary of our best fit model for the velocity distribution of cluster galaxies. {\bf Left panel:} Fraction of dynamically associated and orbiting galaxies as a function of projected cluster radius, as labeled.  {\bf Right panel:} Line-of-sight velocity dispersion for orbiting and infalling galaxies as a function of projected cluster radius, as labelled.}
    \label{fig:model}
\end{figure*}


As before, we use MCMCs to determine the posterior distribution of our model parameters using the total likelihood given by Equation~\ref{full_like}, though now each individual likelihood is given by Equation~\ref{final_li}. Our best fit (maximum likelihood) model is shown in Figure~\ref{fig:histograms}, and provides an excellent description of the data.  Figure~\ref{fig:histograms} vividly illustrates how the fraction of orbiting galaxies decreases with increasing radius, finally disappearing when $R=\rt$. We further note that the fraction of orbiting galaxies at $R\approx \rlambda \approx 0.5\rt$ is $\approx 55\%$, in reasonable agreement with the simulation results of \citet{farahietal16}.  

We can directly compare the constraints from our new model of the velocity distribution of cluster galaxies to the result from section~\ref{two population methods}, and Figure~\ref{fig:fbound} in particular.  The blue bands in Figure~\ref{fig:fbound} correspond to the best fit model for the fraction of dynamically associated galaxies $\fbnd$, and the total velocity dispersion of said galaxies, $\sigmabnd$.  To compute the latter, we rely on the fact that the dynamically associated galaxies are a combination of \bound and infalling galaxies.  From equation~\ref{final_li}, it is clear that the velocity dispersion of dynamically associated galaxies is related to that of \bound and infalling galaxies via
\begin{align}
\sigmabnd^2 = \fvir \sigmavir^2 + (1-\fvir)\sigmainf^2.
\end{align}
We use the above expression, evaluated at the pivot richness and redshift of our sample, to compute the velocity dispersion $\sigmabnd$ from our model.  In Figure~\ref{fig:fbound}, the width of the bands correspond to the 68\% regions as constrained using our MCMCs.  Evidently, our model provides an excellent description of the data.  A graphical summary of our model can be found in Figure~\ref{fig:model}, while the corresponding best fit values for our parameters can be found in Table~\ref{tab:params}.  

One surprising result that is apparent from Table~\ref{tab:params} is the fact that $\alpha_{\rm los} \approx 0.5$.  This trend is detected at high significance due to the large number of galaxies (as indicated by the small error bars at large velocities in Figure~\ref{fig:histograms}), but does little to the qualitative appearance of the fits due to the large value of the pivot line-of-sight velocities (i.e., the line-of-sight velocity component is roughly flat).  As such, it is not obvious to us how to interpret this result.  As we cautioned earlier, however, we do expect selection effects may lead to scalings of the line-of-sight component with cluster richness.


\begin{figure*}
    \centering
    \includegraphics[width=0.49\textwidth]{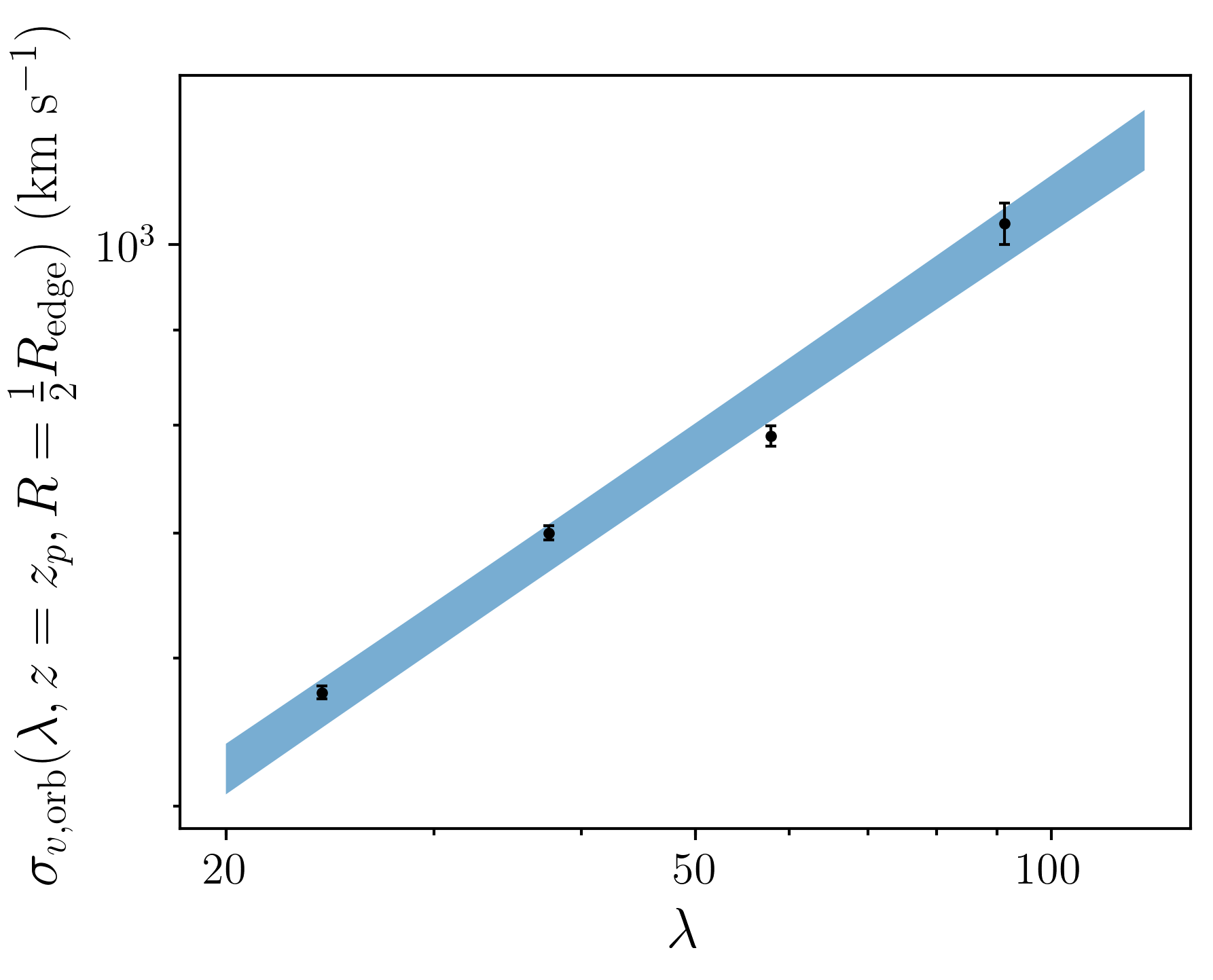}
    \includegraphics[width=0.49\textwidth]{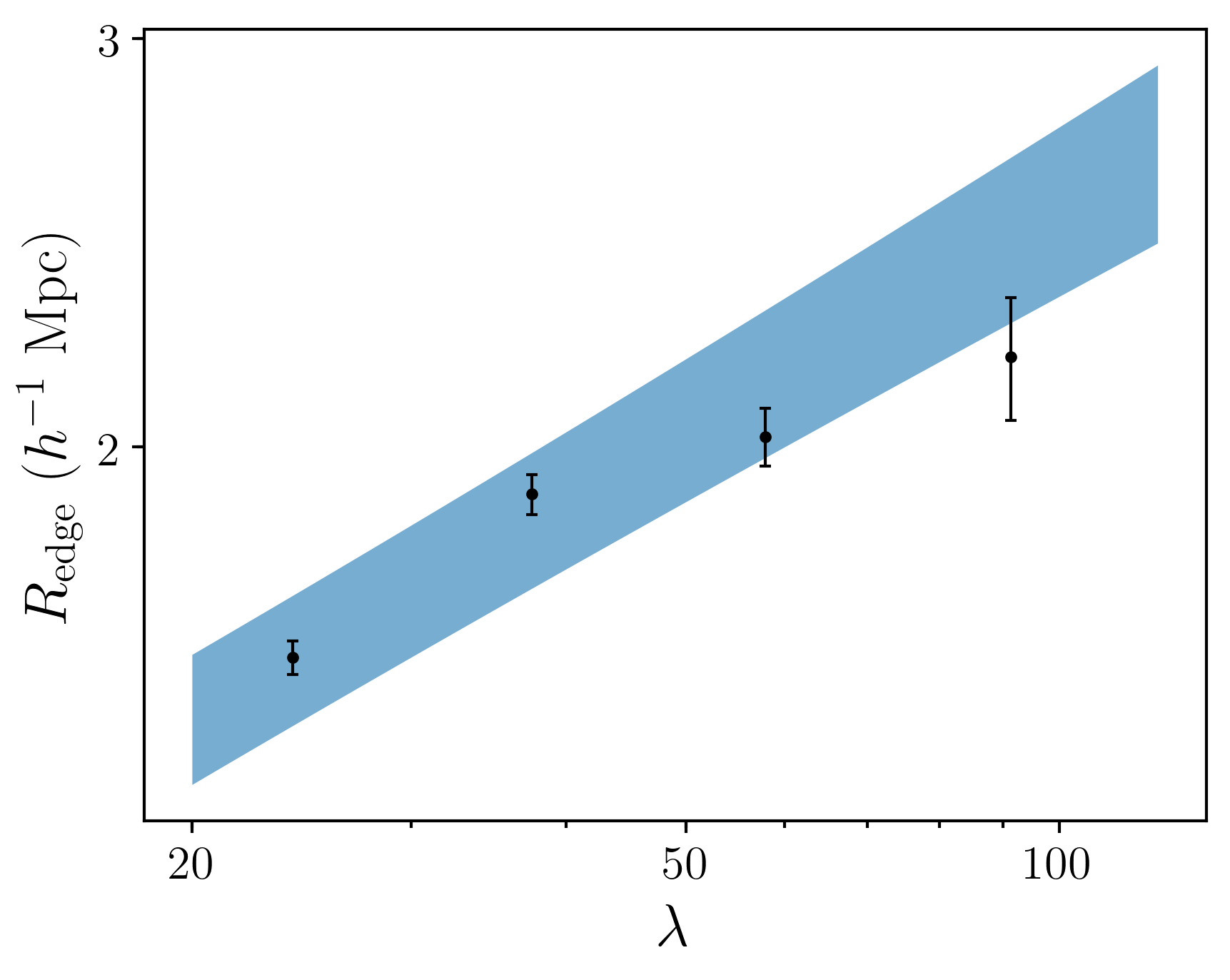}
    \caption{Richness dependence of the orbiting velocity dispersion $\sigmavir$ (left) and edge radius $\rt$ (right). Each data point is the best fit velocity dispersion or edge radius, evaluated at the median richness of each of the four redshift bins we considered.  The small error bars in these measurements reflect the fact that many of the model parameters varied in our full model have been held fixed to their best fit values, as constrained from our global model. The blue bands show the 68\% confidence regions as recovered from our global fit.
    }
    \label{fig:binned}
\end{figure*}


As a consistency check for our analysis, we repeat our measurements, only now we further bin the galaxy clusters as a function of cluster richness.  We split our galaxy clusters into four richness bins, $\lambda \in [20,30),$ $[30,50)$, $[50,80)$, and $[80,120)$.  Unfortunately, our cluster sample is sufficiently sparse that doing so increases the noise in the recovered parameters substantially.  Since our goal here is simply to provide a consistency test, we address this difficulty as follows:
\begin{enumerate}
    \item We assume that the \it ratio \rm of the pivot velocity dispersions $\sigmapinf$ and $\sigmaplos$ relative to the velocity dispersion $\sigmapvir$ of orbiting galaxies is constant.  This reduces the number of free velocity dispersions to one.
    \item We assume that the radial profiles describing the fraction of dynamically associated and orbiting galaxies as a function of cluster radius are fixed at the best fit values. 
    \item We assume that the richness and redshift scaling parameters (i.e. the $\alpha$'s and the $\beta$'s) are fixed at their best fit values.
    \item We assume that the parameter $k$ governing the radial dependent of the velocity dispersion of orbiting galaxies is fixed at its best fit value. 
\end{enumerate}

With these assumptions in hand, we rerun our likelihood for galaxies in individual richness bins, fitting for the velocity dispersion of orbiting galaxies and the edge radius parameters ($\sigmapvir$ and $\rtp$).  The results are shown in Figure~\ref{fig:binned}.  We can see that our richness-binned measurements are in reasonable agreement with the trends inferred from our global fit to the full data set, though there may be some evidence of an increasing slope in the $\rt$--$\lambda$ relation as we move towards low richness systems.  Possible evidence of selection effects preferentially impacting low richness redMaPPer clusters is discussed in \citet{desy1_clusters}.


\section{Summary and Conclusions}

We have measured and characterized the projected phase space distribution of galaxies in the vicinity of SDSS \redmapper\ clusters.  Our main findings are:
\begin{itemize}
    \item The distribution of line-of-sight velocities can be qualitatively described using two components, a roughly Gaussian peak due to galaxies dynamically associated with the cluster, and a ``shelf'' of unassociated galaxies with apparently large velocities (Figure~\ref{fig:histo1}).
    \item The radial dependence of both the fraction of galaxies that belong to the Gaussian peak, and the velocity dispersion of that peak, exhibit a sharp feature at a characteristic radius which we label $\rt$, the edge radius (Figure~\ref{fig:fbound}).  We find $\rt/\rlambda\approx 2.2$.  The velocity dispersion of the Gaussian peak for radii $R>\rt$ is approximately constant.
\end{itemize}

We have argued that the phase space structure seen in Figures~\ref{fig:histo1} and \ref{fig:fbound} can be understood with the following model:
\begin{itemize}
        \item Galaxies near a galaxy cluster come in three ``flavors'': \bound galaxies, infalling galaxies, and random line-of-sight projections.
        \item The line-of-sight ``velocity dispersion'' of the infalling and line-of-sight galaxies is independent of radius, but may scale with richness and redshift as power-laws.
        \item The velocity dispersion of \bound cluster galaxies decreases with increasing radius.
        \item There is an edge radius $\rt$ beyond which there are no \bound galaxies.
        \item Galaxy clusters are self-similar when radially dependent quantities are plotted as a function of $R/\rt$. 
\end{itemize}

Our work is most closely related to the pioneering work by \citet{zuweinberg13}, and the more recent update to that work by \citet{hamabataetal19}.  Both of these works are based on numerical simulations.  The similarities are immediately obvious: these works split galaxies into virialized (their nomenclature) and infalling galaxies, and model the full three-dimensional phase space structure of the galaxies in the vicinity of galaxy clusters as measured in simulations.  Many of the qualitative conclusions of our work are already apparent in these works, though the emphasis and conclusions drawn are different.  In particular, both \citet{zuweinberg13} and \citet{hamabataetal19} were primarily interested in characterizing the infalling region of the galaxy clusters, and the extent to which these infall regions can be used for cluster mass calibration and studies of modified gravity \citep[e.g.][]{zuetal14}.  Consequently, little emphasis was placed on the prominent feature in the velocity data occurring at $R\approx 2\ \hMpc$.  

Our main contribution in this context focuses on the galaxy velocity distribution at ``small'' radii ($R\leq 5\ \hMpc$). We suggest that the sharp feature seen in the velocity distribution of galaxies (Figure~\ref{fig:histo1}) is evidence of a bona fide halo edge, possibly related to the splashback radius.  With this insight in hand, it becomes natural to assert cluster self-similarity in the regions interior to $\rt$, thereby simplifying the original models of \citet{zuweinberg13} and \citet{hamabataetal19}.  On the other hand, our analysis is blind to the infalling velocity profile $\avg{v_r|R}$, which was the driving force behind the \citet{zuweinberg13} and \citet{hamabataetal19} analyses.

Our results provide strong motivation for taking a new look at the phase space of dark matter substructures in numerical simulations.  In a companion paper (Aung et al. 2020), we perform such an analysis, providing theoretical confirmation that the distribution of \bound cluster galaxies have a sharp edge beyond which only infalling galaxies can be found.  Here, ``orbiting'' refers to substructures that have had one pericenter pass in their orbit around the central halo.  Most of these orbiting galaxies appear to be bound in the sense that their velocities are lower than the escape velocity of the halo at their location.  Aung et al. 2020 also demonstrate that the halo edge we have identified is a constant multiple of the splashback radius, where the multiplicative constant is roughly redshift and mass independent.  In a separate work, Garc\'ia et al. 2020 (in preparation) identifies halo edges through detailed modeling of the halo--mass correlation function.  It remains to be seen whether that halo boundary is well matched to the boundary defined by the transition from orbiting to infalling-only galaxy populations.  Should these boundaries match, the case for a the existence of a true halo edge will be the strongest it has ever been.

In a follow-up paper (Aung et al., in prep), we will demonstrate that we can recover the velocity dispersion of various cluster galaxy components from line-of-sight measurements, at least in the absence of selection effects. Future studies need to demonstrate how the measurement of the edge radius is affected by observational effects such as the miscenetring and projection effects, the latter of which plagued the original measurement of the ``splashback'' radius \citep[e.g.,][]{farahietal16,buschwhite17,zuetal17}. 

The presence of the edge radius $\rt$ in the galaxy velocity dispersion, as well as the simplicity of our model, opens the door to multiple follow-up studies. In particular, our models may be used to estimate the velocity dispersion of \bound cluster galaxies, which can in turn be used to estimate cluster masses, all free from contamination by infalling galaxies.  Likewise, the mass dependence of the velocity dispersion of orbiting galaxies and the cluster radius $\rt$ provides a critical consistency check that can enhance mass calibration efforts for cluster cosmology. Moreover, observationally, it is possible to estimate the mass of a galaxy cluster from $\sigmavir$, which can in turn be used to estimate $\rt$.  This makes it possible to perform distance-ladder measurements using the angular scale $\theta_{\rm edge}=\rt/D_A$.  We will investigate this possibility in an upcoming work (Wagoner et al., in preparation).

{\it Acknowledgements:} We acknowledge useful conversations with regard to this work with Susmita Akhikari, Eric Baxter, Chihway Chang, Bhuvnesh Jain and Keiichi Umetsu. ER was supported by the DOE grant DE-SC0015975. DN acknowledges support by National Science Foundation grant AST-1412768 and the facilities and staff of the Yale Center for Research Computing.
ER \& DN also acknowledge funding from the Cottrell Scholar program of the Research Corporation for Science Advancement, which supported PT during the summer of 2019. SS was supported by funding from the Gruber Science Fellowship.

Funding for the Sloan Digital Sky Survey IV has been provided by the Alfred P. Sloan Foundation, the U.S. Department of Energy Office of Science, and the Participating Institutions. SDSS-IV acknowledges
support and resources from the Center for High-Performance Computing at
the University of Utah. The SDSS web site is www.sdss.org.

SDSS-IV is managed by the Astrophysical Research Consortium for the 
Participating Institutions of the SDSS Collaboration including the 
Brazilian Participation Group, the Carnegie Institution for Science, 
Carnegie Mellon University, the Chilean Participation Group, the French Participation Group, Harvard-Smithsonian Center for Astrophysics, 
Instituto de Astrof\'isica de Canarias, The Johns Hopkins University, Kavli Institute for the Physics and Mathematics of the Universe (IPMU) / 
University of Tokyo, the Korean Participation Group, Lawrence Berkeley National Laboratory, 
Leibniz Institut f\"ur Astrophysik Potsdam (AIP),  
Max-Planck-Institut f\"ur Astronomie (MPIA Heidelberg), 
Max-Planck-Institut f\"ur Astrophysik (MPA Garching), 
Max-Planck-Institut f\"ur Extraterrestrische Physik (MPE), 
National Astronomical Observatories of China, New Mexico State University, 
New York University, University of Notre Dame, 
Observat\'ario Nacional / MCTI, The Ohio State University, 
Pennsylvania State University, Shanghai Astronomical Observatory, 
United Kingdom Participation Group,
Universidad Nacional Aut\'onoma de M\'exico, University of Arizona, 
University of Colorado Boulder, University of Oxford, University of Portsmouth, 
University of Utah, University of Virginia, University of Washington, University of Wisconsin, 
Vanderbilt University, and Yale University.
  
\bibliographystyle{mnras}
\bibliography{database.bib} 


\label{lastpage}

\end{document}